# Towards Network Behaviour Trend Evaluation in Software Defined Network (SDN) Considering the number of paths

Alireza Shirmarz[*]
Department of Computer Engineering, Amirkabir University (Polytechnic Tehran), Tehran, Iran.
E-mail: A.shirmarz@aut.ac.ir

## Abstract

There is a wide range of topologies to use in simulation that can make research divergency; therefore, we propose a topology set that can be used in research of network behaviour in Software Defined Network (SDN). This paper can unite the trend research that is doing in different aspects of SDN. One of the most effective items which show the behaviour of the proposed model in SDN is the number of paths that exist between each couple of nodes; hence, we propose three basic topologies to show this parameter. This paper is useful for those who are working on SDN and intend to evaluate the effect of their proposal considering the number of paths. Finally, three topologies called sparse, partial-mesh and full-mesh will be introduced in this paper.

# 1- Introduction

Software Defined Network (SDN) is an architecture that makes the network more programmable, flexible and manageable [1][2][3][4]. This architecture includes three layers and three APIs which are shown in Fig. 1. The key idea in SDN is control and data traffic separation that has caused the network more programmable and flexible as if it has been proposed for the future internet architecture [5][1]. This architecture can be used in Wide Area Network (WAN), Wireless Local Area Network (WLAN), Data Centers (DC), Internet of Things (IoT) and Cloud as mentioned in [4].

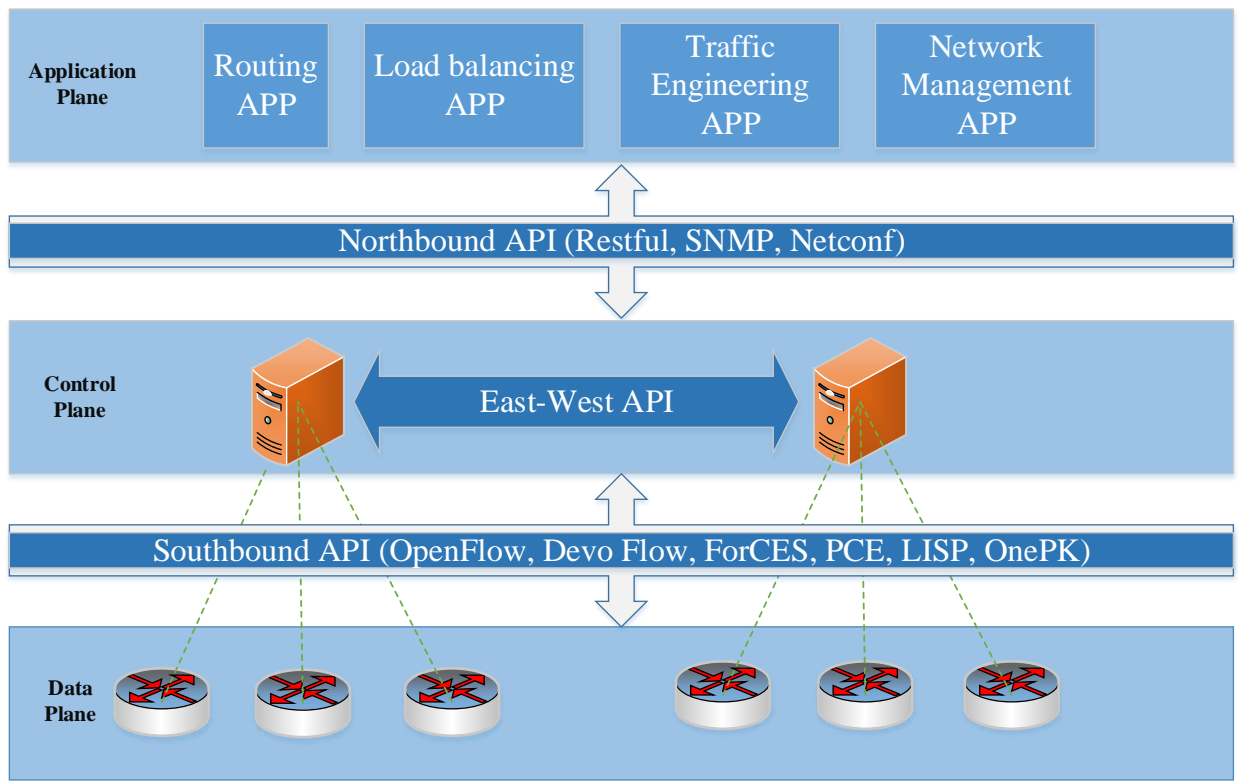

Fig. 1 The global structure of Software Defined Network [4]

One of the most important metrics which has a direct effect on the network behaviour is the number of paths between each couple of nodes. Therefore, a set of topologies are required to show the effect of this item on network behaviour. To unite the research procedure and evaluate it more accurately, we will propose a model to design topologies for network behaviour evaluation. These topologies are sparse, partial-mesh and full-mesh which will be presented in the next section.

## 2- The Proposed Network Behaviour Evaluation Model

To evaluate the proposed model in SDN, we can use Mininet which is an emulator, but there are different topologies and traffic generations models that can make the research results for evaluation complicated. In this section, we present the topology and traffic generation model to help the evaluation of those pieces of research which have been proposed in SDN in order to simplify the evaluation of the results.

### 2.1. Topology

Topology is defined by the number of nodes and the way that these nodes are connected in the simulation.

The number of nodes can be different based on the CPU and RAM capacity on which the simulation should be done.

It can be assumed that topology is a graph consisting of nodes and edges. In this graph, $N$ designates the number of nodes, and $E$ points to the number of edges.

Three conditions can be considered based on the number of connections in SDN, (i) The worst use-case which there is only one way between each couple of nodes and this scenario includes the minimum number of connections, (ii) the ordinary situation has an averaged number of links that increase the number of paths between a couple of nodes compared with the worst topology, but this scenario is not full mesh topology, (iii) the full-mesh topology which contains the maximum number of links as if all nodes are connected to each other.

The worst topology is called 'Sparse Topology', the ordinary scenario is called 'Partial Mesh' and the full-mesh is called the same 'Full-mesh Topology'.

The number of Edges based on the number of nodes is presented below:

a) Sparse Topology → $E = N$

b) Partial-Mesh Topology → $E = \frac{\left(\left[\frac{N\times(N-1)}{2}\right]+N\right)}{2}$

c) Full-Mesh Topology → $E = \frac{N\times(N-1)}{2}$

Full-Mesh topology connects all nods, but Partial-Mesh and Sparse Topology need to select the connected nodes which random algorithm is proposed to choose a couple of nodes to be connected in this proposal.

Each link has a set of features that should be designated with the features including bandwidth, propagation delay, jitter and packet loss rate. These attributes limit the network topology considering the network link features. The steps of this task are presented below:

a) Set Minimum and maximum value by the network experts for bandwidth, delay, jitter and PLR
b) Choose the random value that is limited between min and max values for each feature
c) Set chosen value to each link

Links features are chosen as mentioned in Table 2.

Table 1. Links value designation

| Value Metric | Min | Max | Value Designation Strategy |
|---|---|---|---|
| Bandwidth | a (int) | b (int) | Random(min=a, max=b), int |
| Delay | a' (float) | b'(float) | Random(min=a', max=b'), float |
| Jitter | a''(float) | b''(float) | Random(min=a'', max=b''), float |
| PLR | a'''(float) | b'''(float) | Random(min=a''', max=b'''), float |

## 2.2. Traffic Generation Model

To guide a fair environment to evaluate, four metrics are reckoned to be designated in each SDN simulation. These metrics are:

a) Bandwidth
b) Delay
c) Jitter
d) Packet Loss Rate (PLR)

It is proposed that the maximum and minimum values should be set and then the value should be choose based on a random algorithm. These values present the flow which should be injected from the designated nodes and handled by the controller. The minimum and maximum values should be set by the designers and the designated value will be set randomly as shown in Table 2.

Table 2. Flow metrics value designation

| Value Metric | Min | Max | Value Designation Strategy |
|---|---|---|---|

| Flow Bandwidth Requirement | a (int) | b (int) | Random(min=a, max=b), int |
| --- | --- | --- | --- |
| Flow Delay Requirement | a' (float) | b'(float) | Random(min=a', max=b'), float |
| Flow Jitter Requirement | a''(float) | b''(float) | Random(min=a'', max=b''), float |
| Flow PLR Requirement | a'''(float) | b'''(float) | Random(min=a''', max=b'''), float |

The flows are generated with the features which have been determined in this subsection to evaluate the proposed model fairly.

The source and destination of flows are generated based on the topology that has been designed in sub-section 2-1.

## 3- Dataset Sample

A dataset including 5, 7 and 9 nodes has been generated that can be used as a sample [6]. In this dataset, features related to flows are presented by 'F' and the features related to links are indicated by 'R'.

## 4- Conclusion

In this paper, we have proposed a procedure to model the network for SDN. This model can help those working on SDN and can ease the evaluation. This model focuses on the effects of the number of paths. Therefore, it can guide the research on this topic for researchers all over the world with a proposed guided model for simulation.